\documentclass[aps,prb,twocolumn,superscriptaddress]{revtex4-1}

\usepackage[colorlinks=true, allcolors=blue]{hyperref}
\usepackage{graphicx}
\usepackage{amsmath}
\usepackage{amssymb}
\usepackage{changes}
\usepackage{comment}
\usepackage{braket}
\usepackage{blkarray}
\usepackage{cancel}
\usepackage{soul}

\DeclareGraphicsExtensions{.png,.jpg,.eps}
\usepackage{xcolor}

\newcommand{\cg}[1] {\underline{\textit{#1}}\color{black}\normalsize}

\begin{document}

\author{D. Soriano}
\affiliation{\mbox{Radboud University, Institute for Molecules and Materials, Heyendaalseweg 135, 6525 AJ Nijmegen, The Netherlands}}
\author{A.~N. Rudenko}
\affiliation{\mbox{Radboud University, Institute for Molecules and Materials, Heyendaalseweg 135, 6525 AJ Nijmegen, The Netherlands}}
\affiliation{\mbox{Department of Theoretical Physics and Applied Mathematics, Ural Federal University, 620002 Ekaterinburg, Russia}}
\author{M.~I. Katsnelson}
\affiliation{\mbox{Radboud University, Institute for Molecules and Materials, Heyendaalseweg 135, 6525 AJ Nijmegen, The Netherlands}}
\affiliation{\mbox{Department of Theoretical Physics and Applied Mathematics, Ural Federal University, 620002 Ekaterinburg, Russia}}
\author{M. R\"osner}
\affiliation{\mbox{Radboud University, Institute for Molecules and Materials, Heyendaalseweg 135, 6525 AJ Nijmegen, The Netherlands}}

\title{Environmental Screening and Ligand-Field Effects to Magnetism in CrI$_3$ Monolayer}

\begin{abstract}

    We present a detailed study on the microscopic origin of magnetism in suspended and dielectrically embedded CrI$_3$ monolayer. 
    To this end, we down-fold two distinct minimal generalized Hubbard models with different orbital basis sets from \emph{ab initio} calculations using the constrained random phase approximation. 
    Within mean-field approximation, we show that these models are capable of describing the formation of localized magnetic moments in CrI$_3$ and of reproducing electronic properties of full \emph{ab initio} calculations. 
    We utilize the magnet force theorem to study microscopic magnetic exchange channels between the different orbital manifolds. 
    We find a multi-orbital super-exchange mechanism as the origin of magnetism in CrI$_3$ resulting from a detailed interplay between effective ferro- and anti-ferromagnetic Cr-Cr $d$ coupling channels, which is decisively affected by the ligand (I) $p$ orbitals. 
    We show how environmental screening such as resulting from encapsulation with hexagonal boron nitride (hBN) of the CrI$_3$ monolayer affects the Coulomb interaction in the film and how this successively controls its magnetic properties. 
    Driven by a non-monotonic interplay between nearest and next-nearest neighbour exchange interactions we find the magnon dispersion and the Curie temperature to be non-trivially affected by the environmental dielectric screening. 
    
\end{abstract}

\date{\today}

\maketitle

\section{Introduction}

    Ferromagnetic (FM) layered materials hold high promises for becoming one of the key ingredients in future spintronic nanodevices \cite{KleinJarillo2018,KezilebiekeLiljeroth2020,CardosoRossier2018,LyonsTartakovskii2020} based on van-der-Waals heterostructures. 
    Recent observations of magnons in thin layers of chromium trihalides via inelastic electron tunneling spectroscopy (IETS) \cite{KleinJarillo2018,GhazaryanMisra2018} and magneto-Raman spectroscopy \cite{CenkerXu2020} have also opened new ways to explore magnon-based low-consumption spintronics at the two-dimensional (2D) limit \cite{LiuvanWees2020}. 
    Whether or how the magnetic properties are modified or stabilized in these heterostructures is, however, still under debate.
    Recently reported theoretical predictions suggest, for instance, a strong dependence of the magneto-optical response on the film thickness and the spin-orbit coupling of CrI$_3$ \cite{WuLouie2019}, while recent experimental studies point towards the possibility to tune the magnetic properties of monolayer and bilayer CrI$_3$ using electrostatic gating \cite{SorianoKatsnelson2020,JiangMak2018,JiangShan2018,HuangXu2018}. 
    These observations highlight the importance of addressing how the environment of layered magnetic systems modify their electronic and magnetic properties via proximity, gating, or screening effects.

    For the case of chromium trihalides, CrX$_3$ (X = I, Br Cl), the nearest neighbour magnetic exchange couplings have been theoretically reported to be between $1$ and $3.2\,$meV \cite{LadoRossier2017,ZhangLam2015,KashinRudenko2020,TorelliOlsen2019,BesbesSolovyev2019,XuBellaiche2018} depending on the material and calculation scheme, which renders rigorous theoretical descriptions absolutely necessary. 
    This holds especially with respect to the Coulomb interactions in layered material, which are enhanced due reduced polarization in the surrounding. At the same time changes to the environmental polarization and thus to the environmental screening can significantly modify the Coulomb interaction within the layered material. 
    Thus, properties of layered materials affected by the Coulomb interaction, such as the formation of excitons and plasmons or the stabilization of charge and superconducting order, are in principle strongly dependent on the material's environment allowing for Coulomb engineering of these many-body properties \cite{RosnerWehling2015,RosnerWehling2016,raja_coulomb_2017,steinhoff_exciton_2017,waldecker_rigid_2019,van_loon_coulomb_2020}.
    So far it is not known how efficient Coulomb engineering can be for tailoring the magnetic ground state or magnonic excitations of layered magnetic systems.
    For the case of CrI$_3$ both can be described by effective spin Hamiltonians describing exchange interactions between the Cr atoms as the spin carriers. Thus any changes from environmental screening to the exchange couplings of this effective Hamiltonian can give rise to changes in the magnetic transition temperature or the magnon spectrum opening new routes towards the control and potential tailoring of magnetic properties of chromium trihalides.   

    In the following, we study in detail the impact of Coulomb interactions on the magnetic properties of suspended and dielectrically embedded monolayer CrI$_3$. We start from spin-unpolarized \emph{ab initio} band structure calculations for monolayer CrI$_3$ using density functional theory (DFT), which we use to down-fold two minimal models defined by localized Wannier orbitals describing the Cr $d$ states only and a combined ($d$+$p$)-basis also taking the I $p$ states into account. After constructing the corresponding Coulomb tensors from constrained Random Phase Approximation (cRPA) calculations we solve the resulting multi-orbital Hubbard models within mean-field Hartree-Fock (HF) theory. The resulting interacting and spin-resolved quasi-particle band structures are used afterwards to evaluate the orbitally resolved magnetic exchange interactions using the magnetic force theorem (MFT). Using our Wannier Function Continuum Electrostatics (WFCE) approach we additionally include the environmental screening effects to the Coulomb interactions, which finally allows us to study how the microscopic exchange coupling channels, the magnetic transition temperature, and the magnon dispersions are affected by different parts of the Coulomb tensor and/or by the dielectric encapsulation of the CrI$_3$ monolayer. 

\section{Results and Discussion}
    
    \subsection{Ab initio calculations}
        
    \begin{figure}[htbp]
     \centering
     \includegraphics[width=0.35\textwidth]{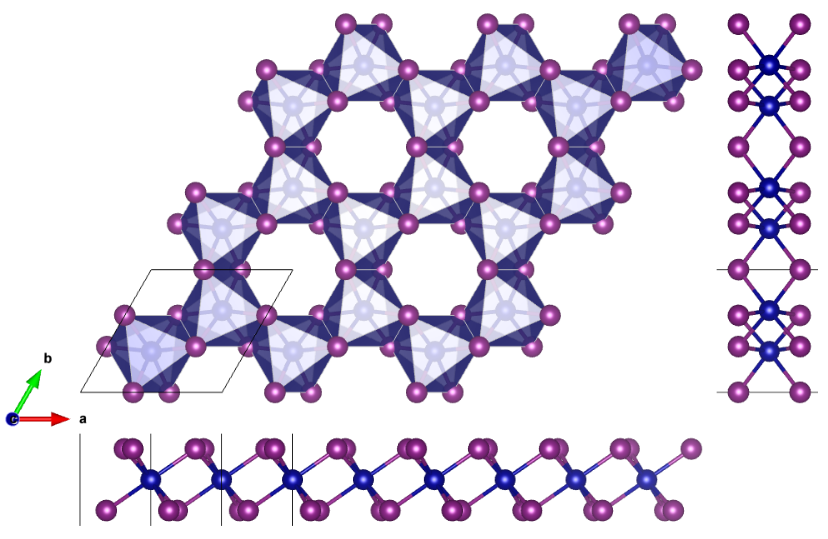}
     \includegraphics[width=0.45\textwidth]{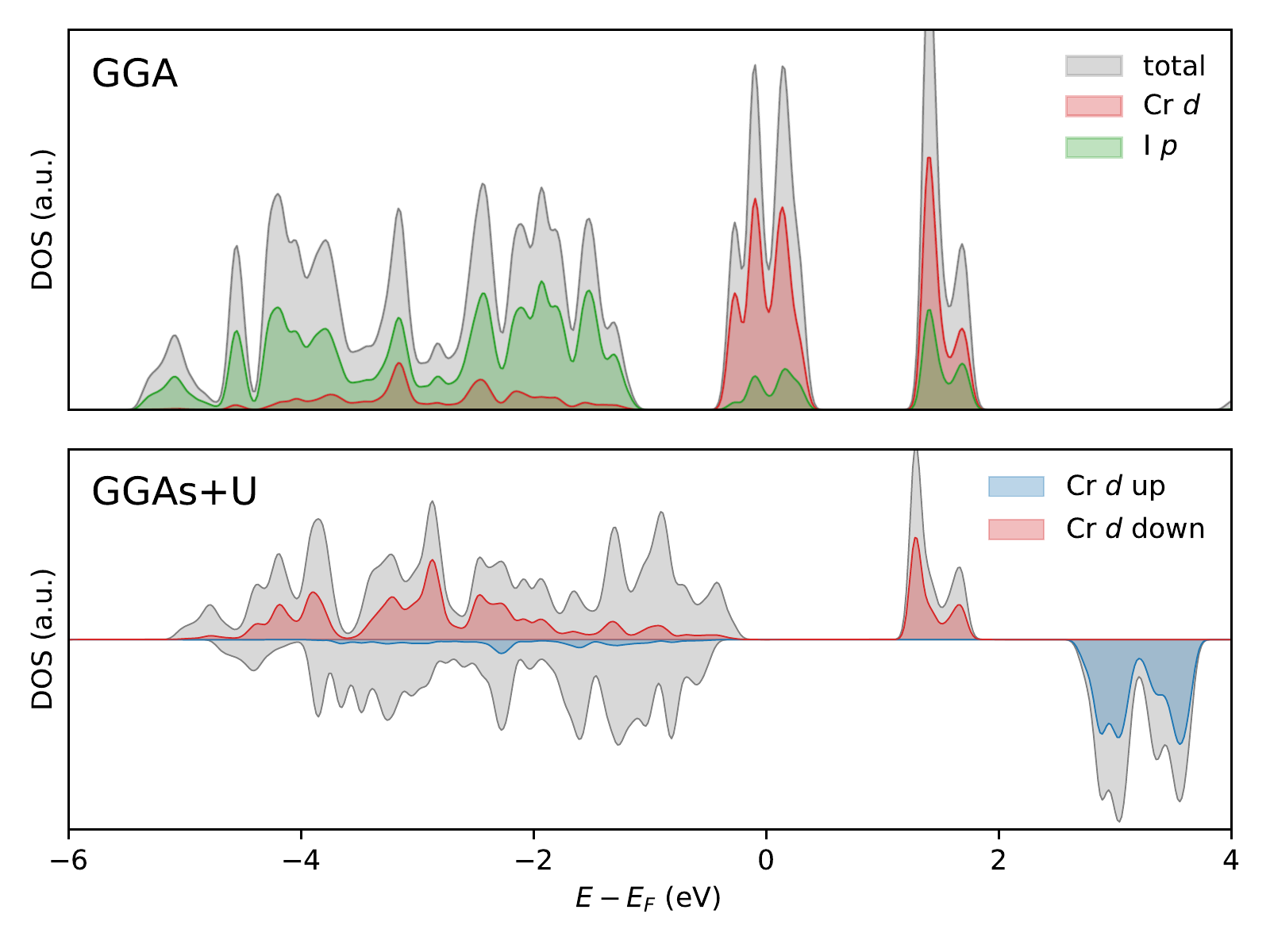}
     \caption{\textbf{CrI$_3$ lattice and ab initio electronic structure.} (a) Lattice structure with primitive unit cell. (b) GGA density of states with I $p$ and Cr $d$ contributions. (c) GGAs+$U$ density of states separated into spin up and down components with Cr $d$ contributions using $U = 4\,$eV and $J = 0.6\,$eV. \label{fig:structure}}
    \end{figure}
    
    In the crystal structure of monolayer CrI$_3$ the Cr atoms are arranged on a honeycomb lattice surrounded by the I ligands forming an edge-sharing octahedral environment around each metal ion, as depicted in Fig.~\ref{fig:structure}~(a). The ligand field splits the Cr $d$-orbitals into two sets, namely, the $t_{2g}$ ($d_{xy},d_{xz},d_{yz}$) and the $e_g$ ($d_{x^2-y^2},d_{z^2}$) \cite{LadoRossier2017}. Spin un-polarized first-principles calculations in the Generalized Gradient Approximation (GGA) level predict a metallic ground state with half-filled bands of predominantly Cr $t_{2g}$ character separated by sizable gaps from fully occupied bands of mostly I $p$ character and unoccupied bands of Cr $e_g$ type, as shown in the orbital resolved density of states in Fig.~\ref{fig:structure}~(b). Spin-polarized calculations in the GGAs+$U$ approximation \cite{ldapu} shift down the majority spin $t_{2g}$ bands which hybridize with the ligand $p$-bands. The unoccupied majority spin $e_g$ bands appear at approx. $1.5\,$eV above the valence states followed by the minority spin $t_{2g}$ and $e_g$ bands, respectively, resulting in a FM insulating state, as depicted in Fig.~\ref{fig:structure}~(c).
    Although LSDA+$U$ approaches seem to reasonably describe the magnetic properties of CrI$_3$, its validity still needs to be benchmarked with higher-level theories such as dynamical mean field theory, as the local Coulomb repulsion $U$ is rather large compared to the $t_{2g}$ band width as discussed in detail later. Thus, to fully microscopically understand each part of the problem, ranging from hybrdizations effects, to the Coulomb tensor, or the choice of basis set we proceed in the following with a down-folding procedure to generate reliable material-realistic minimal models.
    
    \subsection{Model Hamiltonian}
        
    In the following we will construct model Hamiltonians of the form
    \begin{equation}
        \mathcal{H} = \mathcal{H}^{W} + \mathcal{H}^{U} - \mathcal{H}^{DC}. \label{eq:Hamiltonian}
    \end{equation}
    with $\mathcal{H}^{W}$ describing the non-interacting kinetic terms and $\mathcal{H}^{U}$ describing the local Coulomb interactions of the Cr $d$ orbitals. Depending on the chosen basis we additionally use a double-counting correction $\mathcal{H}^{DC}$. The non-interacting part of the Hamiltonian is defined by long-range hopping matrix elements $t_{mn}$ and reads
    \begin{eqnarray}
        \mathcal{H}^{W} & = & \sum_{k,m,n,\sigma} t_{mn}(k) d^{\dagger}_{m\sigma}(k) d_{n\sigma}(k),
    \end{eqnarray}
    while the Coulomb term takes local density-density $U_{mm'}$ and Hund's exchange $J^H_{mm'}$ interactions on the Cr atoms into account and is defined by
    \begin{align}
        \mathcal{H}^{U} = &\sum_{m,\sigma}{U_{mm}\hat{n}_{m\sigma}\hat{n}_{m\bar{\sigma}}} \\ 
        + &\sum_{m\neq m',\sigma}{U_{mm'}\hat{n}_{m\sigma}\hat{n}_{m'\bar{\sigma}}} \notag \\
        + &\sum_{m\neq m', \sigma}{(U_{mm'}-J^H_{mm'})\hat{n}_{m\sigma}\hat{n}_{m'\sigma}} \notag \\
        - &\sum_{m\neq m', \sigma}{(J^H_{mm'})d^{\dagger}_{m\sigma}d_{m\bar{\sigma}}d^{\dagger}_{m'\bar{\sigma}}d_{m'\sigma}}, \notag
    \end{align}
    where $\hat{n}_\sigma = d^\dagger_{m\sigma}d_{m\sigma}$ is the number operator and $\sigma$ is the spin index. These Hamiltonians are solved within a mean-field HF ansatz and analyzed in terms of its microscopic magnetic properties as described in the Methods section.
    
    \subsection{Basis Sets and Non-Interacting Hamiltonians}

        Motivated by the orbital-resolved \emph{ab initio} non-magnetic density of states depicted in Fig.~\ref{fig:structure}~(b), we utilize two minimal basis sets: one including only effective Cr $d$ Wannier orbitals and one also including I $p$ states, which are referred to as $d$-only and ($d$+$p$) model, respectively, in the following. To this end, we start with conventional DFT calculations utilizing the Perdew–Burke–Ernzerhof GGA exchange correlation functional \cite{PhysRevLett.77.3865} within a PAW basis \cite{paw,paw2} as implemented in the \emph {Vienna Ab initio Simulation Package} ({\sc vasp}) \cite{KRESSE199615,PhysRevB.54.11169} for CrI$_3$ monolayers with a lattice constant of $a_0 \approx 6.97\,$\AA\ embedded in a supercell with a height of $35\,$\AA. We use ($16 \times 16 \times 1$) $k$-meshes together with an energy cutoff of $230\,$eV and apply a Methfessel-Paxton smearing of $\sigma=0.02\,$eV. For the $d$-only model we project the Kohn-Sham states onto Cr-localized $d$ orbitals (with a rotated axis parallel to the tetragonal main axis) and maximally localize them afterwards using the {\sc wannier90} package \cite{MOSTOFI2008685}. For the ($d$+$p$)-model, we start with the same Cr-centered $d$ orbitals and add also I-centered $p$ orbitals to the list of initial projects. In this case we do not perform a maximal localization since it would increase the localization of the I $p$ orbitals at the cost of a delocalization of Cr $d$ orbitals. The resulting localized Wannier orbitals are used in a subsequent step to calculate the needed hopping matrix elements for the definition of $\mathcal{H}^{W}$. The resulting Wannier models perfectly interpolate the mostly Cr $d$ Kohn-Sham states in the $d$-only model as well as also the I $p$ states in the ($d$+$p$)-model.

    \subsection{Constrained Random Phase Approximation}
    
        The Coulomb interaction matrix elements of $\mathcal{H}^{U}$ are evaluated using the Cr $d$ Wannier orbitals from these two models within the cRPA scheme \cite{aryasetiawan_frequency-dependent_2004} according to
        \begin{align}
            U_{ijkl} = \braket{ w_i w_j | \mathcal{U} | w_k w_l }.
        \end{align}
        Here $\cal{U}$ represents the partially screened Coulomb interaction defined by
        \begin{align}
            \mathcal{U} = \mathcal{V} + \mathcal{V} \, \pi_\text{rest} \,\mathcal{U},
        \end{align}
        with $\mathcal{V}$ being the bare Coulomb interaction and $\pi_\text{rest}$ the partial or rest-polarization from all states expect those of the correlated sub-space defined by the Cr $d$ states. ${\pi}_\text{rest}$ thus describes screening from all other Cr as well as all I states including a significant amount of empty states from the full Kohn-Sham basis. In detail, we use in total $128$ bands and define ${\pi}_\text{rest}$ by explicitly excluding all Kohn-Sham states between $-0.5$ and $3\,$eV from the full polarization. To this end, we use a recent cRPA implementation by Kaltak available in {\sc vasp} \cite{KaltakcRPA}. From this we can extract the full 4th order Coulomb tensor in the basis of the correlated Cr $d$ states. Within $\mathcal{H}^{U}$ we restrict ourselves however to local density-density elements $U_{mm'} = U_{mm'm'm}$ and local Hund's exchange elements $J^H_{mm'} = U_{mm'mm'}$ with $m$ and $m'$ labeling $d$ orbitals on the same Cr atom.  
        
    \subsection{Minimal Cr $d$-only basis}
        
    \begin{table}
        \centering
        \caption{Bare ($v$/$j^H$) and cRPA screened ($U$/$J^H$) local Coulomb interaction matrix elements from the $d$-only basis set. Density-density matrix elements are listed on the upper triangle and Hund's exchange elements are underlined \label{tab:UJdOnly}}
        \begin{ruledtabular}
            \begin{tabular}{l|lll|ll}
            $j^H / v$ & $t_{2g}$ & $t_{2g}$ & $t_{2g}$ & $e_{g}$ & $e_{g}$ \\ \hline 
            $t_{2g}$ & 15.501	& 14.384	& 14.383	& 13.086	& 12.783 \\
            $t_{2g}$ & \cg{0.535}	& 15.506	& 14.386	& 13.086	& 12.786 \\
            $t_{2g}$ & \cg{0.535}	& \cg{0.535} 	& 15.505	& 12.635	& 13.236 \\  \hline 
            $e_{g}$  & \cg{0.318} 	& \cg{0.318} 	& \cg{0.515} 	& 12.313	 & 11.288 \\
            $e_{g}$  & \cg{0.449}	& \cg{0.449} 	& \cg{0.252} 	& \cg{0.512} & 12.312 \\ \hline\hline
            $J^H / U$ & $t_{2g}$ & $t_{2g}$ & $t_{2g}$ & $e_{g}$ & $e_{g}$ \\ \hline 
            $t_{2g}$ & 3.546 &	2.531 &	2.531 &	2.669 &	2.419 \\
            $t_{2g}$ & \cg{0.487} &	3.547 &	2.532 &	2.668 &	2.421 \\
            $t_{2g}$ & \cg{0.487} &	\cg{0.487} & 3.547 &	2.296 &	2.793 \\ \hline
            $e_{g}$  & \cg{0.301} &	\cg{0.301} & \cg{0.465} & 3.108 &	2.224 \\
            $e_{g}$  & \cg{0.410} &	\cg{0.410} & \cg{0.247} & \cg{0.442} &	3.108 \\
            \end{tabular}
        \end{ruledtabular}
    \end{table}
    
    We start with analyzing the bare and cRPA screened density-density Coulomb matrix elements as obtained from the $d$-only basis. The full matrices are shown in Tab.~\ref{tab:UJdOnly} together with the corresponding Hund's exchange elements. In all cases the resulting density-density matrices are approximately of the form
    \begin{align}
        &v_{mm'} \approx \label{eq:vmn} \\
        &\begin{pmatrix}
            v_t & v_t - 2j^H_t & v_t - 2j^H_t & v_{et} - 2j^H_1 & v_{et} - 2j^H_2 \\
            v_t - 2j^H_t & v_t & v_t - 2j^H_t & v_{et} - 2j^H_1 & v_{et} - 2j^H_2 \\
            v_t - 2j^H_t & v_t - 2j^H_t & v_t & v_{et} - 2j^H_3 & v_{et} - 2j^H_4 \\
            v_{et} - 2j^H_1 & v_{et} - 2j^H_1 & v_{et} - 2j^H_3 & v_e & v_e - 2j^H_e \\
            v_{et} - 2j^H_2 & v_{et} - 2j^H_2 & v_{et} - 2j^H_4 & v_e - 2j^H_e & v_e \\
        \end{pmatrix}, \notag
    \end{align}
    where $v_t$ ($j^H_t$) and $v_e$ ($j^H_e$) are intra-orbital density-density (inter-orbital Hund's exchange) matrix elements within the $t_{2g}$ and $e_g$ manifolds, respectively, $v_{et} = 1/2 ( v_e + v_t)$, and $j^H_{1 \dots 4}$ are inter-orbital Hund's exchange elements between the two manifolds. This form of the density-density Coulomb matrix is similar to the one obtained for a fully rotational-invariant $d$ shell which is here, however, perturbed due to the ligand-induced crystal-filed splitting. Therefore, instead of five ($U_0$, $J^H_{1 \dots 4}$) or even just three ($U_0$, $F_2$, $F_4$ \cite{slater_theory_1929}) parameters to represent the full density-density matrix, we need here eight. As a result, the $t_{2g}$ and $e_g$ channels themselves are easily parameterized using two Hubbard-Kanamori parameters ($v$ and $j^H$). The inter-channel elements show, however, a significant orbital dependence which can be represented by the four $j^H_{1\dots4}$ Hund's exchange elements.
    
    Since the Wannier orbital spread $\Omega_\alpha = \braket{w_\alpha|r^2|w_\alpha} - \braket{w_\alpha|r|w_\alpha}^2$ \cite{MOSTOFI2008685} of the $t_{2g}$ wave functions ($\Omega_{t_{2g}} \approx 3\,$\AA$^2$) is smaller then the corresponding $e_g$ one ($\Omega_{eg} \approx 5.3\,$\AA$^2$) the bare intra-orbital density-density elements $v_t \approx 15.5\,$eV are larger than the $v_e \approx 12.3\,$eV elements and also represent the largest elements in $v_{mm'}$. This is also reflected in the intra-channel $j^H_t \approx 0.54\,$eV and $j^H_e \approx 0.51\,$eV elements which are, however, still rather similar.
    The inter-channel $j^H_{1\dots4}$ vary between about $0.25\,$eV and $0.47\,$eV.
    We note that these bare Hund's exchange interactions are significantly smaller than the approximated values in bulk Cr of $J^H \approx 0.75\,$eV \cite{SasiogluBluegel2011}, which we attribute here to the enhanced Wannier function spread of the effective Cr $d$-only basis.

    The cRPA density-density matrix elements $U_{mm'}$ are significantly reduced by factors between $1/4$ to nearly $1/6$ by the rest-space screening from the other Cr and I states. Although this effective screening is strongly orbital dependent, it does not change the overall orbital structure of the Coulomb matrix depicted in Eq.~(\ref{eq:vmn}). 
    We find $U_t \approx 3.6\,$eV and $U_e \approx 3.1\,$eV, which are still rather large compared to the band-width of the half-filled Cr $t_{2g}$ band of about $1\,$eV. Thus even taking cRPA screening into account correlation effects can be expected to play an important role.
    The screened Hund's exchange interactions, $J^H_t \approx 0.49\,$eV and $J^H_e \approx 0.44\,$eV, are reduced by no more than $10\%$ in comparison to the bare values, which also holds for the inter-channel ones.
    
\begin{figure}
        \centering
        \includegraphics[width=0.45\textwidth]{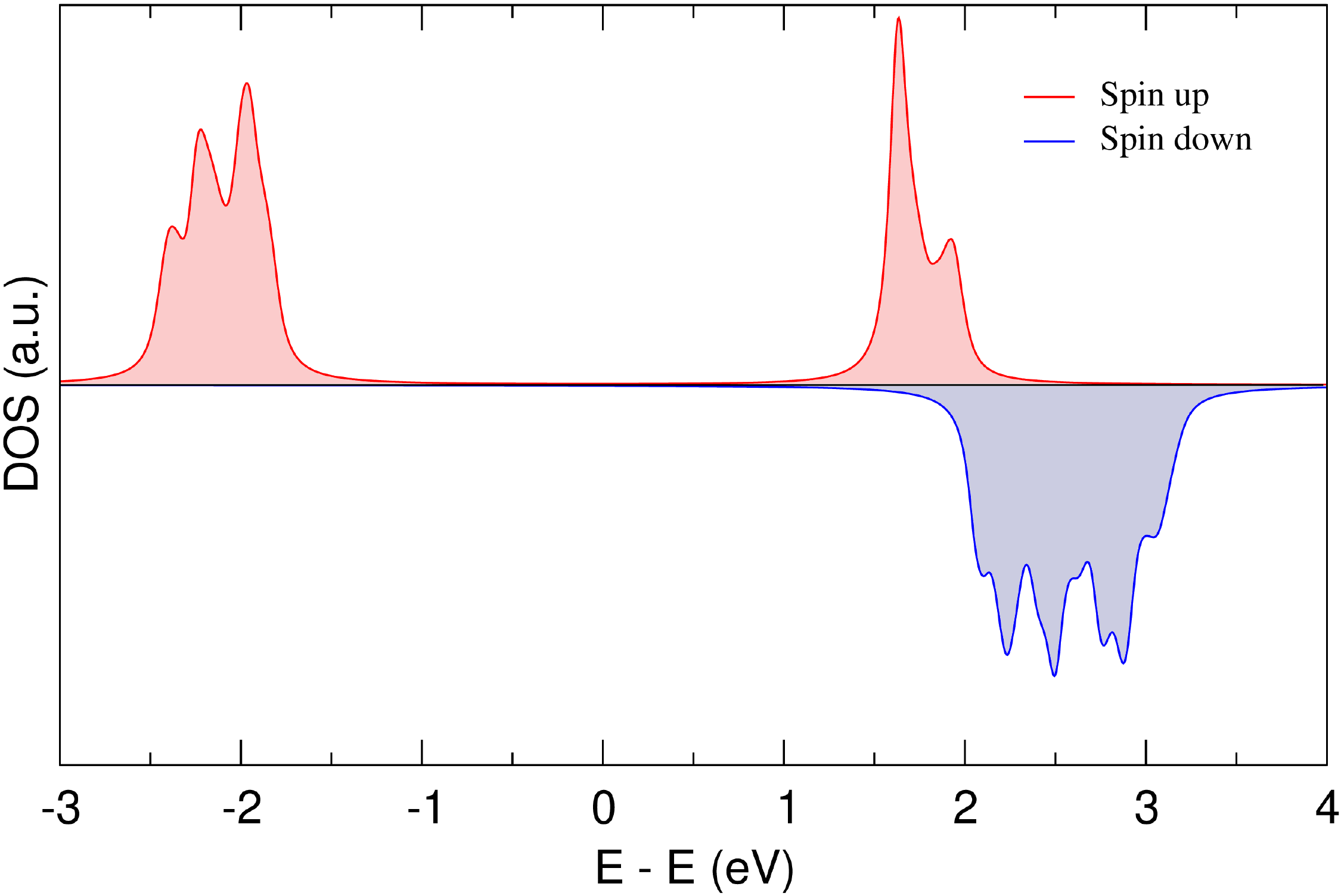}
        \caption{\textbf{$d$-only density of states.} Red and blue denote spin up and down, respectively. Coulomb matrix elements from Tab.~\ref{tab:UJdOnly} are used.}
        \label{fig:WanHF-d-Bands}
    \end{figure}

    We proceed with analyzing how these local cRPA-screened Coulomb interactions affect the band structure of the Cr $d$-only model within mean-field theory. 
    As depicted in Fig.~\ref{fig:WanHF-d-Bands} the Coulomb interactions drive the systems into an insulator with a sizeable band gap and the same band ordering as known from LSDA(+$U$) calculations (i.e. fully occupied $t_{2g}$ followed by completely empty $e_{g,\uparrow}$, $t_{2g,\downarrow}$, and $e_{g,\downarrow}$ manifolds). On a qualitative level this minimal basis thus seems to be capable of reproducing the Cr $d$ band structure of full \emph{ab initio} calculations.
    To analyze to what extend this minimal model is also capable of reproducing the super-exchange mechanism responsible for the FM ordering in CrI$_3$ (see Methods) we have applied the MFT to these HF solutions in a subsequent step (for details see Methods). The resulting orbitally-resolved exchange couplings are given in Tab.~\ref{tab:Exchange} and show a strong \emph{AFM} total coupling of $J = -1.964$ meV for nearest neighbours and $J' = -0.042$ meV for next nearest neighbours, driven by a strong AFM $J_\text{t2g-t2g}$ interaction. This is obviously in contradictions to LSDA(+$U$) calculations and to all available experimental data. This wrong prediction can be understood from the Kugel-Khomskii (KK) formalism \cite{KashinRudenko2020,kugel_jahn-teller_1982}, which describes the total magnetic exchange as the sum of the $t_{2g}$-$t_{2g}$ and $t_{2g}$-$e_g$ contributions as $J(r) = J_\text{t2g-t2g}(r) + J_\text{eg-eg}(r)$ with
    \begin{eqnarray}
        J_\text{t2g-t2g}(r) &\approx& -\frac{2 \tilde{t}_\text{t2g-t2g}^2(r) }{\tilde{U}} \notag \\
        J_\text{eg-eg}(r) &\approx& \frac{2 \tilde{t}_\text{t2g-eg}^2(r) \tilde{J}^H}{(\tilde{U} + \Delta) (\tilde{U} + \Delta - \tilde{J}^H)}. \notag
    \end{eqnarray}
    Here, $\tilde{t}_\text{t2g-t2g}(r)$ and $\tilde{t}_\text{t2g-eg}(r)$ are effective hopping matrix elements between the different orbital channels, $\Delta$ represents the crystal field splitting, and $\tilde{U}$ and $\tilde{J}^H$ are the averaged Coulomb and Hund's exchange parameters. We immediately understand that $J_\text{t2g-t2g}$ is by definition of AFM nature and controlled only by the effective $t_{2g}$-$t_{2g}$ hopping and the density-density channel of the Coulomb interaction ($\tilde{U}$). In contrast, the FM channel, $J_\text{eg-eg}$, is controlled by the effective $t_{2g}$-$e_g$ hopping and both $\tilde{U}$ and $J_H$. By using the hopping values of our Wannier construction, given in Tab.~\ref{tab:hoppings} of the Supplement, and the averaged electron-electron interactions from the cRPA Coulomb tensor ($\tilde{U}\approx3.4$ and $\tilde{J}_H\approx0.4$ eV), we obtain an AFM intralayer exchange coupling in line with the MFT results.  
    \begin{table}
        \centering
        \caption{Orbital-resolved exchange interactions (in meV) from MFT. \label{tab:Exchange}}
        \begin{ruledtabular}
            \begin{tabular}{lcccccc}
                & $J_\text{t2g-t2g}$ & $J_\text{t2g-eg}$ & $J$ & $J'_\text{t2g-t2g}$ & $J'_\text{t2g-eg}$ & $J'$ \\
                d only &  $-2.472$ & $+0.508$ & $-1.964$ & $-0.137$ & $+0.095$ & $-0.042$ \\
                d+p & $-0.864$ & $+2.620$ & $+1.756$ & $-0.046$ & $+0.398$ & $ +0.352$
            \end{tabular}
        \end{ruledtabular}    
    \end{table}
    We carefully checked that this wrong KK prediction also holds in the case of the extended ($d$+$p$) basis by using effective hoppings from this models with and without $p$-mediated hopping, as summarized in Tab.~\ref{tab:hoppings} of the Supplement. In both cases, using only the direct hoppings as well as those after integrating our the I $p$ contributions, we find that the effective n.n. $\tilde{t}_\text{t2g-t2g}$ is significantly larger than the $\tilde{t}_\text{t2g-eg}$ one so that the AFM coupling channels always dominates. Thus the correct microscopic origin of the FM coupling in CrI$_3$ cannot be described in a Cr $d$-only basis and can neither be modelled within the KK approach.
  
    \subsection{Extended ($d$+$p$)-basis}
     
    \begin{table}
    \centering
            \caption{Bare ($v$/$j^H$) and cRPA screened ($U$/$J^H$) local Coulomb interaction matrix elements from the ($d$+$p$) basis set. Density-density matrix elements are listed on the upper triangle and Hund's exchange elements are underlined \label{tab:UJd+p}}
            \begin{ruledtabular}
                \begin{tabular}{l|lll|ll}
                $j / v$ & $t_{2g}$ & $t_{2g}$ & $t_{2g}$ & $e_{g}$ & $e_{g}$ \\ \hline
                $t_{2g}$ & 18.480 &	17.138 &	17.138 &	18.172 &	17.739 \\
                $t_{2g}$ & \cg{0.664} &	18.479 &	17.138 &	18.172 &	17.738 \\
                $t_{2g}$ & \cg{0.664} &	\cg{0.664} &	18.479 &	17.520 &	18.390 \\ \hline
                $e_{g}$  & \cg{0.485} &	\cg{0.485} &	\cg{0.804}  &	19.875 &	18.159 \\
                $e_{g}$  & \cg{0.697} &	\cg{0.697} &	\cg{0.378}  &	\cg{0.858}  &	19.877 \\ \hline \hline
                $J^H / U$ & $t_{2g}$ & $t_{2g}$ & $t_{2g}$ & $e_{g}$ & $e_{g}$ \\ \hline 
                $t_{2g}$ & 3.993 &	2.769 &	2.769 &	3.146 &	2.794 \\
                $t_{2g}$ & \cg{0.600} &	3.993 &	2.769 &	3.146 &	2.794 \\
                $t_{2g}$ & \cg{0.600} &	\cg{0.600} &	3.994 &	2.618 &	3.323 \\ \hline
                $e_{g}$  & \cg{0.460} &	\cg{0.460} &	\cg{0.714} &	4.175 &	2.646 \\
                $e_{g}$  & \cg{0.629} &	\cg{0.629} &	\cg{0.375} &	\cg{0.765} &	4.176 \\
                \end{tabular}
            \end{ruledtabular}
    \end{table}
        
    Motivated by the wrong predictions of the minimal Cr $d$-only model, we expanded the basis to also include the ligand $p$ contributions. This ($d$+$p$)-basis has a significant impact to the local Cr Coulomb matrix elements as summarized in Tab.~\ref{tab:UJd+p}. Due to the presence of the I $p$ orbitals the Cr $d$ Wannier functions are more localized ($\Omega \approx 1\,$\AA$^2$) so that the resulting bare matrix elements are significantly enhanced as compared to the $d$-only case. Also the density-density matrix elements in the $e_g$ channel are now larger than in the $t_{2g}$ channel. The cRPA screening due to all other Cr and all I states, again significantly reduce all matrix elements in an orbital-dependent manner. The final intra-orbital density-density interactions are now on the order of $4\,$eV with inter-orbital elements of the order of $3\,$eV. Notably, the cRPA screened Hund's exchange interactions are also enhanced by the increased localization of the Cr $d$ states. They are, however, still on the order of $0.5$ to $0.7\,$eV and thus still significantly smaller than the common approximation of $0.9\,$eV.
    
    As before in the $d$-only basis the interaction term $\mathcal{H}^U$ of our Hamiltonian from Eq.~(\ref{eq:Hamiltonian}) acts on the Cr $d$ states only. The kinetic part, $\mathcal{H}^W$, however now also includes I $p$ contributions. Thus, to counter-act double counting effects we subtract here the double counting potential defined in Eq.~(\ref{eq:DC}) from the Cr $d$ states using a nominal Cr$^{3+}$ $d$ occupation of $N_{imp}^\sigma = 3$. This corrects for the relative positioning of the interacting Cr $d$ bands with respect to the uncorrelated I $p$ states.
    
    \begin{figure}
        \centering
        \includegraphics[width=0.45\textwidth]{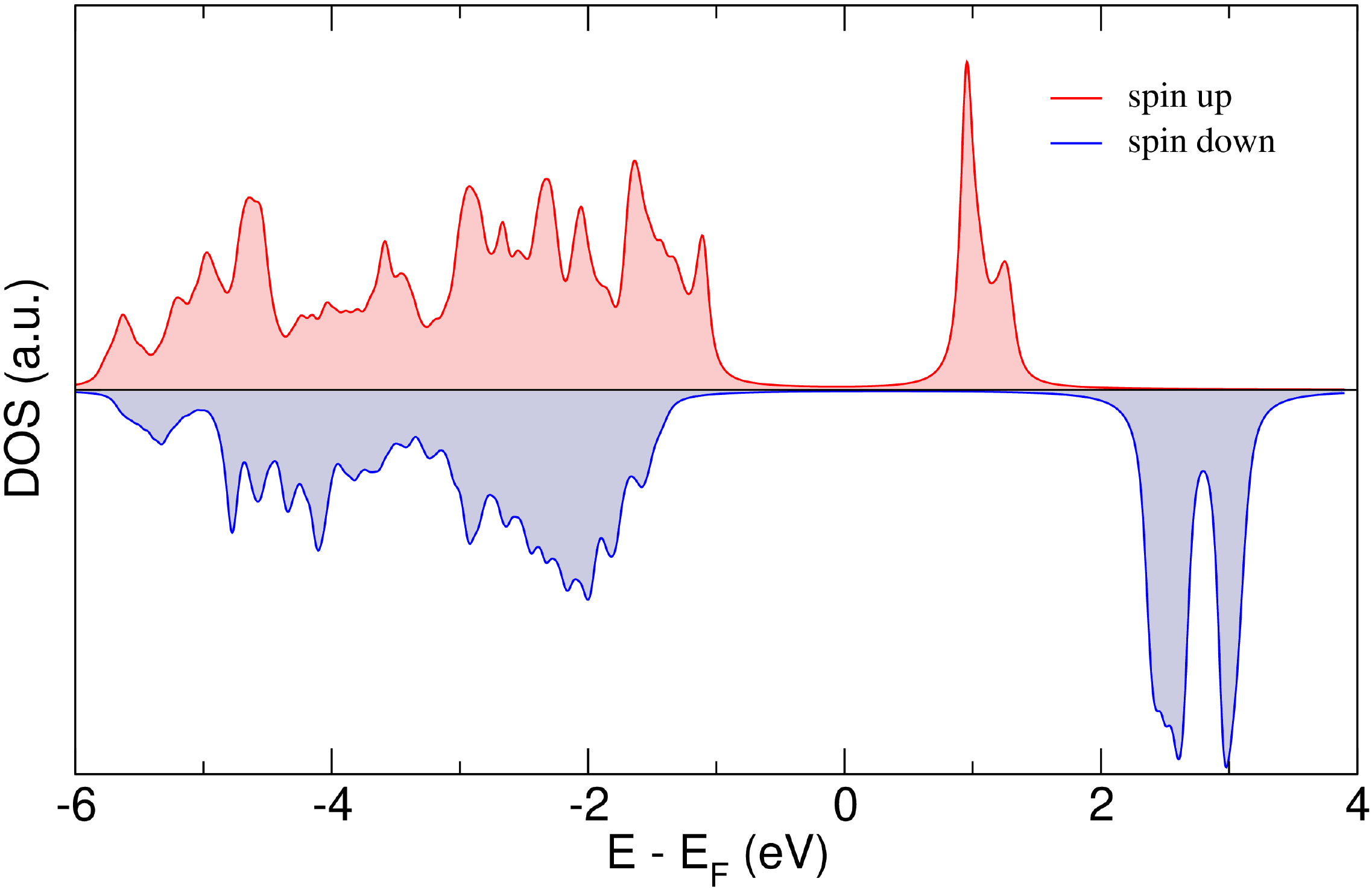}
        \caption{\textbf{Extended ($d$+$p$)-model density of states.} Red and blue denote spin up and down, respectively. Coulomb matrix elements from Tab.~\ref{tab:UJd+p} are used.}
        \label{fig:WanHF-dp-Bands}
    \end{figure} 
    The resulting HF density of states is shown in Fig.~\ref{fig:WanHF-dp-Bands}. 
    In contrast to the $d$-only model we now find a band structure which is vastly reminiscent of the full \emph{ab initio} GGAs+$U$ results shown in Fig.~\ref{fig:structure}.
    Next to the spin-splitting and -ordering also the full and sub band gaps are in good agreement with GGAs+$U$ predictions.
    
    The orbital-resolved intralayer exchange couplings calculated using the MFT are given in Tab.~\ref{tab:Exchange}.
    In contrast to the $d$-only model, the ($d$+$p$)-model gives the correct FM exchange coupling $J \approx 1.76\,$meV ($J' \approx 0.35\,$meV), similar to previously reported values from MFT calculations based on LSDA+$U$ input \cite{KashinRudenko2020}.  
    \begin{figure}[h]
        \centering
        \includegraphics[width=0.45\textwidth]{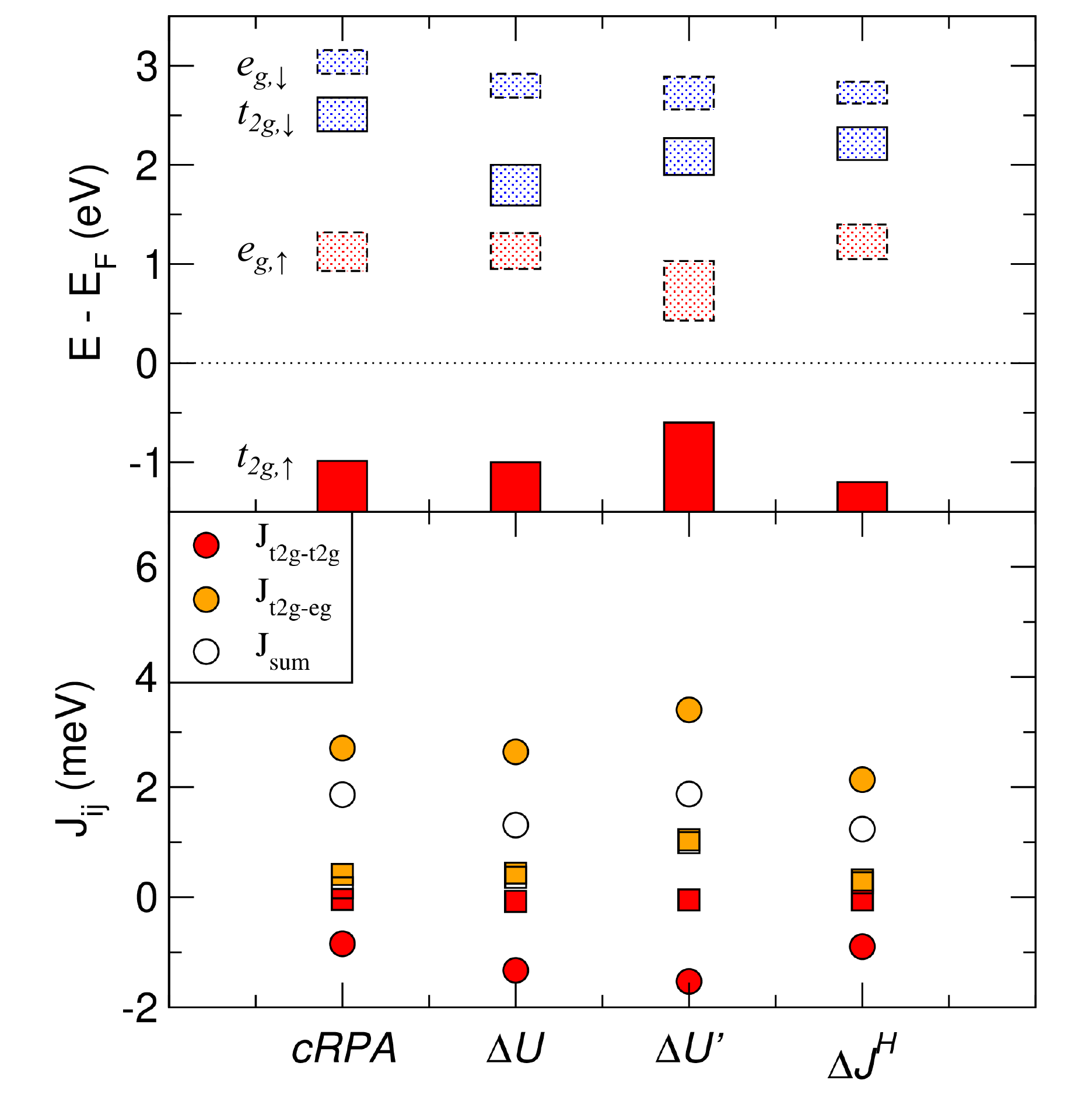}
        \caption{\textbf{Coulomb interaction effects to the electronic and magnetic properties of the ($d$+$p$)-model.} (a) Electronic properties: red and blue stands for spin up and down polarized bands. Continuous and dashed lines correspond to $t_{2g}$ and $e_g$ bands, respectively. Dotted line indicates the Fermi energy position. The size of the boxes is related to the bandwidth. (b) Magnetic Exchange interactions: circles and squares denote 1st and 2nd neighbour intra-layer exchange couplings obtained with the MFT. Red, orange and empty circles stand for $J_\text{t2g-t2g}$, $J_\text{t2g-eg}$, and $J_{sum}$ which is the sum of both interactions. For both, 1st and 2nd neighbours, $t_{2g}$-$t_{2g}$ interactions are always AFM, while the $t_{2g}$-$e_g$ ones are FM. }
        \label{fig:par-effects}
    \end{figure}   
    To microscopically explore how the different Coulomb interaction channels ($U$, $U'$, $J^H$) affect the electronic band structure and the magnetic properties, we present in Fig.~\ref{fig:par-effects} the band widths and positions together with the orbitally-resolved exchange interactions upon individual  \emph{reductions} of each Coulomb channel by $20\%$. Reducing the intra-orbital Coulomb interactions ($\Delta U$) results in a lower spin splitting of the $t_{2g}$ and $e_g$ bands, similar to the single-orbital Hubbard model. The other two cases, namely, $\Delta U'$ and $\Delta J^H$, are more difficult to understand, since these terms mix different types of orbitals. When reducing $U'$, we observe a strong reduction of the splitting between the $t_{2g,\uparrow}/e_{g,\uparrow}$ manifolds, accompanied by a strong gap reduction of about $1\,$eV. In contrast, the splitting between the $t_{2g,\downarrow}/e_{g,\downarrow}$ manifolds is weakly affected by $U'$ compared to the initial full cRPA case. 
    Finally, when we reduce Hund's exchange $J^H$, the $t_{2g,\sigma}/t_{2g,\bar{\sigma}}$ and the $e_{g,\sigma}/e_{g,\bar{\sigma}}$ splittings are both reduced, while the splitting between the majority spin occupied $t_{2g,\uparrow}$ bands and empty $e_{g,\uparrow}$ bands increases with respect to the unperturbed case, leading to a small increase in the bandgap.
    
    These band-structure renormalizations have non-trivial effects to the microscopic exchange interactions, which we depict in Fig.~\ref{fig:par-effects}~(bottom panel) and which can be just partially understood within the KK formalism. In line with KK a reduction of $U$ or $U'$ yields an enhancement of the AFM $J_\text{t2g-t2g}$ channel, as shown in the $\Delta U$ and $\Delta U'$ columns of Fig.~\ref{fig:par-effects}~(bottom panel), while modifications to $J^H$ do not affect $J_\text{t2g-t2g}$ at all. Also in qualitative agreement with KK we find that a reduction of $J^H$ reduces the FM $J_\text{t2g-eg}$. In contrast to the simple model predictions from KK we find from MFT that $J_\text{t2g-t2g}$ is slightly reduced upon reducing $U$ while modifications to $U'$ yields no modifications. This underlines the need for the full microscopic MFT in combination with an extended basis to quantitatively understand magnetism in CrI$_3$. We expect that just an extended model combining the super-exchange mechanism from the Goodenough-Kanamori description and the multi-orbital KK mechanism might yield a qualitatively understanding in line with our MFT findings, which will be studied separately. Here, we conclude that the extended ($d$+$p$)-basis together with the corresponding cRPA Coulomb matrix elements and using a HF solver result in a reliable band-structure and microscopically correct magnetic properties.
    
    \subsection{Substrate Tunability} 
    
\begin{figure*}
        \centering
        \includegraphics[width=0.9\textwidth]{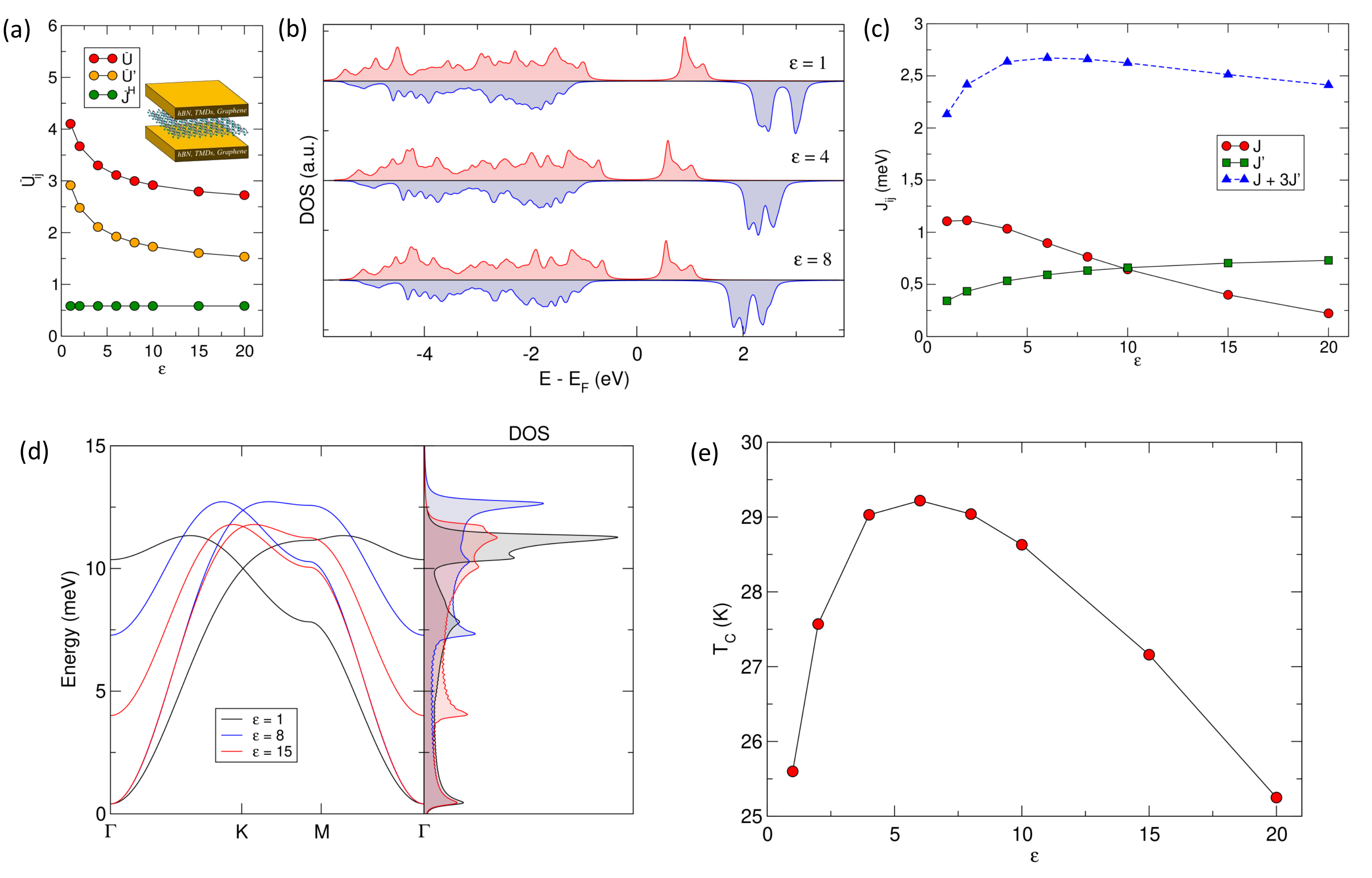}
        \caption{\textbf{Substrate screening effects.} (a) Local Coulomb interactions: red, yellow, and green dots correspond to orbital-averaged intra- ($U$) and interorbital ($U'$) Coulomb and Hund's exchange ($J^H$) interactions. The inset sketches an encapsulated monolayer CrI$_3$. (b) Electronic density of states for $\epsilon = 1$, $\epsilon = 4$, and $\epsilon = 8$. (d) Nearest and next-nearest neighbor magnetic exchange interactions $J$ and $J'$, respectively. The function $J+3J'$ is shown for reference (see text for details). (e) Curie temperature.
        \label{fig:substrate}}
    \end{figure*}

    In the following we proceed with the analysis of dielectric screening effects to the magnetic properties of CrI$_3$ monolayer. We investigate how a dielectric encapsulation, as depicted in the inset of Fig.~\ref{fig:substrate}~(a), modifies the Coulomb interactions in CrI$_3$ and how this affects its bands structure and eventually its microscopic magnetic properties.

    Fig.~\ref{fig:substrate}~(a) summarizes how the orbital averaged local intra- and inter-orbital density-density as well as Hund's exchange matrix elements, scale upon increasing the environmental screening ($\epsilon$).
    Since $\epsilon$ affects the macroscopic monopole-like interactions only, $U$ and $U'$ are equally reduced by $\epsilon$, while $J^H$ is not modified at all. The resulting screening-induced effects will thus be a combination of the $\Delta U$ and $\Delta U'$ columns from Fig.~\ref{fig:par-effects}. 
    
    In Fig.~\ref{fig:substrate}~(b) we show the HF density of states (DOS) for three different values of $\epsilon$. As discussed in the previous section, a decreasing $U$ results in a reduction of the $d$-bands splitting $\Delta_{t_{2g,\sigma}/t_{2g,\sigma}}$ and $\Delta_{e_{g,\sigma}/e_{g,\sigma}}$ and a decrease of the inter-orbital Coulomb terms $U'$ reduces the gap between $t_{2g,\uparrow}$ and $e_{g,\uparrow}$. The overall effect of increasing $\epsilon$ is thus to decrease all gaps between all sub-bands.
    
    \begin{table}[]
        \centering
        \caption{Orbital-resolved nearest- and next-nearest neighbor exchange interactions (in meV) in CrI$_3$ monolayer calculated for different substrate dielectric screening constants. \label{tab:dielectric-orbital}}
        \begin{ruledtabular}
            \begin{tabular}{c|lll|lll}
                $\varepsilon$ & 
                $J_\text{t2g-t2g}$ & 
                $J_\text{t2g-eg}$ & 
                $J$ & 
                $J'_\text{t2g-t2g}$ & 
                $J'_\text{t2g-eg}$ & 
                $J'$ \\ \hline
                1 & $-$0.864 & 2.620 & 1.756 & $-$0.046 & 0.398 & 0.352\\
                2 & $-$1.116 & 2.956 & 1.840 & $-$0.056 & 0.502 & 0.446\\
                4 & $-$1.452 & 3.198 & 1.746 & $-$0.070 & 0.616 & 0.546\\
                6 & $-$1.694 & 3.280 & 1.586 & $-$0.080 & 0.684 & 0.604\\
                8 & $-$1.882 & 3.308 & 1.426 & $-$0.086 & 0.732 & 0.646\\
                10 & $-$2.032 & 3.314 & 1.282 & $-$0.094 & 0.766 & 0.672\\
                15 & $-$2.308 & 3.298 & 0.990 & $-$0.106 & 0.822 & 0.716\\
                20 & $-$2.496 & 3.270 & 0.774 & $-$0.114 & 0.856 & 0.742\\
            \end{tabular}
        \end{ruledtabular}  
        \label{tab:dielectric-orbital}
    \end{table}
    
    In Fig.~\ref{fig:substrate}~(c) we summarize the resulting effects to the magnetic properties.
    Upon increasing $\epsilon$ we find the total nearest-neighbour magnetic exchange decreasing from about $J \approx 1.1\,$meV at $\epsilon=1$ to $J \approx 0.2\,$meV at $\epsilon=20$, while the total next-nearest-neighbour exchange interaction slightly increases from $J' \approx 0.3\,$meV to $J' \approx 0.7\,$meV.
    From Tab.~\ref{tab:dielectric-orbital} we understand that the decreasing trend in $J$ is mostly driven by the enhancement of the AFM coupling within the $J_\text{t2g-t2g}$ channel, while the FM $J_\text{t2g-eg}$ channel is barely affected.
    The next-nearest-neighbour $J'$ is affected simultaneously by both, increasing FM and AFM microscopic exchange interactions. The FM channel grows, however, slightly faster so that the overall FM $J'$ is enhanced.
    From this we expect non-trivial effects of the environmental dielectric screening to the magnetic properties of CrI$_3$ monolayer relevant for most experimental setups dealing with supported and/or encapsulated films.
    
    We start with the analysis of the magnon dispersion which reacts to the environmental screening differently in different parts of the Brillouin zone, as shown in Fig.~\ref{fig:substrate}~(d).
    At the $\Gamma$ point the optical (high energy) branch is continuously lowered in energy upon increasing the screening, while the Dirac point at $K$ is non-monotonously affected. 
    Upon increasing $\epsilon$ from $1$ to $8$ the magnon energy at $K$ first increases before it starts to decrease for larger $\epsilon$.
    This behaviour becomes clear from the spin-wave Hamiltonian Eq.~(\ref{HP-equation}) evaluated at $K$ yielding the degenerated magnon-energies $E(K)=3S(J + 3J' + \lambda)$. 
    As one can see from Fig.~\ref{fig:substrate}~(c), $J+3J'$ is indeed a non-monotonic function of $\epsilon$ with a maximum around $\epsilon=6$.
    For the van-Hove singularities of both magnon branches at the $M$ point we find similar non-trivial and non-monotonic behaviours with $\epsilon$ yielding a partially extended flat dispersion of the optical branch for intermediate $\epsilon$.
    These non-trivial modifications to the magnon dispersion due to changes in $\epsilon$ are also reflected in the total magnon spectrum. Thereby we can relate each (partial) maximum in the magnon spectrum with either the $\Gamma$ or the $M$ point. As a result, these most prominent spectral features either monotonously decrease in energy or follow the non-monotonous trends from the $M$ point.
    
    In Figure Fig.~\ref{fig:substrate}~(e), we additionally show the Curie temperature ($T_C$) for the same $\epsilon$. 
    Again, we find a non-monotonic behavior with an initially increasing $T_C$ upon increasing the screening, a maximal $T_C$ around $\epsilon=6$, and a subsequently strongly decreasing trend.
    This trend approximately follows the spectral peak arising from the optical magnon branch at the $M$ point. 
    Due to its similarity with the $K$-point behaviour we conclude that the initial increasing trend in $T_C$ is driven by the increasing trend of $J'$ while the final decreasing trend is driven by $J$.
    The non-monotonic behavior of $T_C$ is thus due to the non-monotonic interplay between nearest and next-nearest neighbor exchange interactions as a function of the environmental screening.

\section{Conclusions \& Outlook}

    By combining state-of-the-art cRPA-based \emph{ab initio} down-folding with our WFCE approach and the MFT method, we were able to study on a microscopic level how magnetism in CrI$_3$ monolayer builds up and how it is controlled by environmental screening properties. We showed that a mean-field description within the HF approximation to treat local Cr Coulomb interactions is sufficient to reproduce all characteristics of full \emph{ab initio} GGAs+$U$ calculations. From a thorough investigation of different minimal models we understood that only an I $p$-based super-exchange mechanism together with the full multi-orbital $t_{2g}$-$e_g$ structure of the Cr atoms allows for a realistic description of CrI$_3$ magnetism. A minimal model thus needs to involve all Cr $d$ and I $p$ states with sizable Coulomb interactions acting on the Cr $d$ states. We also showed how environmental screening significantly reduces the local Coulomb interactions, which decisively modifies the electronic band structure and which finally yields non-monotonic changes to the microscopic magnetic exchange interactions. In detail we found that dielectric encapsulation of the CrI$_3$ monolayer strongly reduces the nearest-neighbour exchange interaction, while the next-nearest-neighbour interaction is just slightly enhanced, which leaves remarkable footprints in the magnon spectral function and the Curie temperature.
    
    These findings point to a variety of new questions and problems to tackle in the future: On the modeling side we found sizeable local Coulomb interactions as compared to the non-interacting band width, possibly rendering dynamical mean field theory or similar approaches necessary to capture all correlation effects \cite{kvashnin_dynamical_2020,craco_mott_2020,yekta_strength_2021}.
    Our extended ($d$+$p$)-model is an optimal starting ground for studies like these.
    Together with sizeable magnon-phonon couplings \cite{tian_magneto-elastic_2016}, light-matter interactions, and long-range Coulomb interaction effects \cite{WuLouie2019,ke_electron_2021}, we can expect a plethora of correlation effects in this material to be found in the future.
    
    In light of our findings on the environmental-screening-induced modifications to the magnetic properties of monolayer CrI$_3$ we expect that magnetism in multilayer CrI$_3$ is more involving than expected. Next to electronic inter-layer hybridization effects layer-dependent changes to Coulomb interactions seem to be important as well to gain a full understanding. 
    Also the role of anisotropic magnetic interactions, including both symmetric and asymmetric forms, needs to be studied in further detail.
    Substrate effects might be especially relevant in the context of Dzyaloshinskii-Moriya interactions (DMI), as they could be considerably enhanced by breaking/lowering inversion symmetries. At the same time, DMI appear to be promising in for the stabilization of topological magnons \cite{Kvashnin2020} and skyrmionic phases \cite{Behera2019,Behera2019comment} in chromium trihalides. Coulomb engineering of DMI could be considered using a computational scheme similar to the one proposed in Ref.~\onlinecite{Katsnelson2010}.

    Finally, our findings render CrI$_3$ monolayer based heterostructures with spatially structured environments  a possibly fascinating fundamentally new playground to non-invasively structure the magnetic properties of layered magnetic materials similar to what has been discussed for correlation effects in layered semiconductors \cite{RosnerWehling2016,raja_coulomb_2017,utama_dielectric-defined_2019,waldecker_rigid_2019,steinke_coulomb-engineered_2020}.
    Together with the recent discovery of other 2D ferromagnets and antiferromagnets we thus expect that magnetic van der Waals heterostructures are most promising platforms to engineer and design next-generation magnetic and opto-magnetic devices.
    The encapsulation-mediated tunability of the magnetic exchange has important implications in the future application of 2D ferromagnets as spintronic devices. The possibility to combine strong and weak FM regions on 2D ferromagnets using different substrates may find application as memory storage devices. Also, real-space manipulation of the magnon dispersion can open new possibilities for low-consumption magnon-based devices. 
    
\section{Methods}

   \subsection{Hartree-Fock Solver and Double Counting Corrections}
        
        We approximately solve the Hamiltonian from Eq.~(\ref{eq:Hamiltonian}) utilizing a variational single-particle wave function which allows for the breaking of the spin-symmetry. The variational energy is computed by decoupling the interaction terms in the conventional way. The resulting intra-atomic HF Hamiltonian takes the form of an effective single-particle one with local occupations $n_i$ which need to be self-consistently evaluated. It can be divided into spin-conserving ($h^{\uparrow\uparrow}$, $h^{\downarrow\downarrow}$) and spin-mixing terms ($h^{\uparrow\downarrow}$, $h^{\downarrow\uparrow}$). In a general form, the full effective single-particle Hamiltonian reads as:
        \begin{equation}
        \mathcal{H} = \left[
        \begin{array}{cc|cc}
        {h^{\uparrow\uparrow}_{ii}} & {h^{\uparrow\uparrow}_{ij}} & {h^{\uparrow\downarrow}_{ii}} & {h^{\uparrow\downarrow}_{ij}} \\ 
        {h^{\uparrow\uparrow}_{ji}} & {h^{\uparrow\uparrow}_{jj}} & {h^{\uparrow\downarrow}_{ji}} & {h^{\uparrow\downarrow}_{jj}} \\ \cline {1-4}
        {h^{\downarrow\uparrow}_{ii}} & {h^{\downarrow\uparrow}_{ij}} & {h^{\downarrow\downarrow}_{ii}} & {h^{\downarrow\downarrow}_{ij}} \\ 
        {h^{\downarrow\uparrow}_{ji}} & {h^{\downarrow\uparrow}_{jj}} & {h^{\downarrow\downarrow}_{ji}} & {h^{\downarrow\downarrow}_{jj}}
        \end{array}\right],
        \end{equation}
        where the $i,j$ are orbital indices. The explicit expressions for each of the terms in the Hamiltonian are given below:
        \begin{eqnarray}
        h_{ii}^{\sigma\sigma} & = & U_{ii}\rho_{ii}^{\bar{\sigma}\bar{\sigma}} + \sum_{j \neq i} \left(U'_{ij}\rho_{jj}^{\bar{\sigma}\bar{\sigma}} + (U'_{ij}-J^H_{ij})\rho_{jj}^{\sigma\sigma}\right), \nonumber \\
        h_{ij}^{\sigma\sigma} & = & -\sum_{j \neq i}\left((U'_{ij}-J^H_{ij})\rho_{ji}^{\sigma\sigma} + J^H_{ij}\rho_{ji}^{\bar{\sigma}\bar{\sigma}} \right), \nonumber \\
        h_{ii}^{\sigma\bar{\sigma}} & = & - U_{ii}\rho_{ii}^{\bar{\sigma}\sigma} - \sum_{j \neq i} J^H_{ij}\rho_{jj}^{\bar{\sigma}\sigma}, \nonumber \\
        h_{ij}^{\sigma\bar{\sigma}} & = & - \sum_{j \neq i} U'_{ij}\rho_{ji}^{\bar{\sigma}\sigma} ,\nonumber
        \end{eqnarray}
        where $\rho$ is the self-consistent density matrix containing the local occupations: $\rho_{ii}^{\sigma\sigma}=\hat{n}_i^\sigma$ and $\rho_{ij}^{\sigma\bar{\sigma}} = \langle d_{i,\sigma}^\dagger d_{j,\bar{\sigma}} \rangle$.
  
        In order to minimize double-counting errors in the ($d$+$p$)-model due to interaction effects in the hopping matrix elements from the \emph{ab initio} calculations, we subtract a double-counting potential based on the fully localized limit (FLL) \cite{ldapu} approximation:
        \begin{align}
            \mu^{AMF}_{DC} = \bar{U}\left(N_{imp}-\frac{1}{2}\right) - \bar{J}\left(N^\sigma_{imp}-\frac{1}{2}\right) \label{eq:DC}
        \end{align}
        where $\bar{U} = 1/(2l+1)\sum_i U_{ii}$ and $\bar{J} = 1/(4l^2+2l)\sum_{i \neq j} J_{ij}$ are the mean Coulomb and Hund exchange interactions obtained from the cRPA tensors, $N_{imp}$ is the total occupancy of the Cr $d$-orbitals, and $N^\sigma_{imp}$ is the occupancy per spin ($N^\sigma_{imp} = N_{imp}/2$ in the paramagnetic ground state). 
       
     \subsection{Magnetic Force Theorem}
        
        Magnetism in CrI$_3$ and related compounds results from a detailed interplay between local and non-local kinetic and local Coulomb interactions terms yielding an effective magnetic exchange between neighbouring Cr atoms. Generally speaking it can be understood as a super-exchange mechanism mediated by the ligand atoms which follows approximately the Goodenough-Kanamori mechanism \cite{LadoRossier2017}. With an approximate $90^\circ$ angle between neighbouring Cr-I-Cr atoms we thus expect a FM coupling. On a fully microscopic level, Kashin \emph{et al.} \cite{KashinRudenko2020} have recently shown that the total FM exchange interaction results from an interplay between an AFM coupling channel between Cr t$_{2g}$ orbitals and a FM channel between $t_{2g}$ and $e_g$ orbitals. In both, LSDA and LSDA+$U$ calculations this interplay is dominated by the FM $t_{2g}$-$e_g$ channel so that the total exchange interaction is also FM. To understand how this microscopic picture is affected by different choices of the target space, the different orbital channels of the Coulomb tensor, and by environmental screening effects, we follow Kashin \emph{et al.} \cite{KashinRudenko2020} and analyze the results of our HF calculation by means of the MFT \cite{LiechtensteinGubanov1987}, which allows us to calculate the orbitally resolved exchange interaction matrix elements 
        via the second variations of the total energy with respect to infinitesimal rotations of the magnetic moments, leading to the expression \cite{Rudenko2013,Logemann2017}
        \begin{equation}
        \label{eq:Exchange}
        J_{ij}=\frac{1}{2\pi S^2} \int_{-\infty}^{E_F} d\varepsilon  \sum_{\alpha \beta \gamma \delta}  \Im \left[ \Delta_i^{\alpha \beta} G_{ij}^{\beta \gamma 
        \downarrow}(\varepsilon)\Delta_j^{\gamma \delta}G_{ji}^{\delta \alpha \uparrow}(\varepsilon) \right].
        \end{equation}
        Here, Latin (Greek) indices denote atomic (orbital) indices, respectively, $E_F$ is the Fermi energy, and $\Delta_i^{\alpha\beta}=H^{
        \alpha\beta\uparrow}_{{ii}} - H^{\alpha\beta\downarrow}_{{ii}}$ is the exchange splitting matrix defined in the orbital space.
        In Eq.~(\ref{eq:Exchange}), $G_{ij}^{\alpha \beta \sigma}(\varepsilon)= \sum_{\bf k}G_{\bf k}^{\alpha \beta \sigma}(\varepsilon) e^{i {\bf k} \cdot {\bf R}_{ij}}$
        is the real-space Green's function, whose {\bf k}-space matrix representation reads
        \begin {equation}
        \mathcal{G}_{\bf k}^{\sigma}(\varepsilon)=\left[ \varepsilon \mathcal{I} - \mathcal{H}^{\sigma}_{\bf k} + i\eta\mathcal{I} \right]^{-1}, 
        \end{equation}
        where $\mathcal{I}$ is the unity matrix, $\eta \rightarrow 0^+$ is a numerical smearing parameter, ${\bf R}_{ij}$ is the translation vector connecting atoms $i$ and $j$, and $\mathcal{H}^{\sigma}_{\bf k}$ is the reciprocal-space Hamiltonian for spin $\sigma=\uparrow,\downarrow$, whose matrix elements are obtained in the basis of Wannier functions from our HF calculations. Positive and negative $J_{ij}$ correspond here to FM and AFM couplings, respectively.

    \subsection{Magnetic Properties}

        The spin Hamiltonian describing the exchange interactions between Cr atoms in CrI$_3$ can be written as:
        \begin{align}
            \mathcal{H}_{spin} &=\sum_i A(S_i^z)^2 + \frac{1}{2}\sum_{i,j} J_{ij}{\bf S}_i{\bf S}_j \label{spinH-equation}\\
                               &+ \frac{1}{2}\sum_{i,j} \lambda_{ij} S_i^zS_j^z \notag
        \end{align}
        where the first term $A$ describes single-ion anisotropy, $J_{ij}$ is the isotropic Heisenberg exchange, and $\lambda_{ij}$ is the anisotropic symmetric exchange. 
        To calculate the spin-wave spectrum, we transform Eq.~(\ref{spinH-equation}) into a bosonic Hamiltonian using the linearized Holstein-Primakoff transformation with $A = 0$ and $\lambda = 0.09$ meV (see Ref.~\onlinecite{LadoRossier2017}):
        \begin{align}
            \mathcal{H}_{SW} &= \sum_i [3S(J+\lambda) + 6SJ']b^\dagger_ib_i \notag \label{HP-equation} \\
                             &- JS\sum_{\langle ij \rangle}b^\dagger_ib_j 
                              - J'S\sum_{\langle\langle ij \rangle\rangle}b^\dagger_ib_j
        \end{align}
        where $J$ and $J'$ correspond to nearest and next-nearest neighbor isotropic exchange couplings, and $\lambda$ is the nearest neighbor anisotropic symmetric exchange. 
        In {\bf k}-space, the Hamiltonian Eq.~(\ref{HP-equation}) for a honeycomb lattice takes the form
        \begin{equation}
            {\cal H}_{SW}({\bf q})=S
            \begin{pmatrix}
             \epsilon_0 - J'f_2({\bf q})& -Jf_1^*({\bf q}) \\
                -Jf_1({\bf q})    & \epsilon_0 - J'f_2({\bf q})
            \end{pmatrix},
        \end{equation}
        where $\epsilon_0=3J + 6J' + 3\lambda$, and $f_1({\bf q})=\sum_{\bf R}e^{-i{\bf q}\cdot {\bf R}}$ and $f_2({\bf q})=\sum_{\bf R'}e^{-i{\bf q}\cdot {\bf R'}}$ are form factors with ${\bf R}$ and ${\bf R}'$ running over cells with nearest and next-nearest neighbor atoms, respectively.
                The diagonalization of this Hamiltonian yields the magnon spectrum, which reads
                \begin{equation}
            E({\bf q}) = S(3J + 6J' + 3\lambda - J'f_2({\bf q}) \pm J|f_1(\bf q)|). 
        \end{equation}

        To calculate the magnetic $T_C$, we use the Tyablikov's decoupling approximation (also known as RPA) \cite{Rusz} through the expression 
        \begin{align}
            T_C = \frac{1}{2} \frac{S(S+1)}{3k_B} \left(\frac{1}{N}\sum_{{\bf q}} \frac{1}{E({\bf q})}\right)^{-1}.
            \label{Tc-equation}
        \end{align}

    \subsection{Substrate Screening}
    
        Next to the CrI$_3$ intrinsic properties we aim to also understand the role of external screening properties such as resulting for substrate materials or capping dielectrics. To this end, we utilize our WFCE approach \cite{RosnerWehling2015}, which realistically modifies the CrI$_3$ Coulomb interaction tensor according to dielectric environmental screening. In this way we will be able to understand how the electronic band structure and the microscopic magnetic properties are affected, e.g., by encapsulating CrI$_3$ with hBN or under the influence of bulk substrates. 
        
\begin{figure}[htbp]
         \centering
         \includegraphics[width=0.49\textwidth]{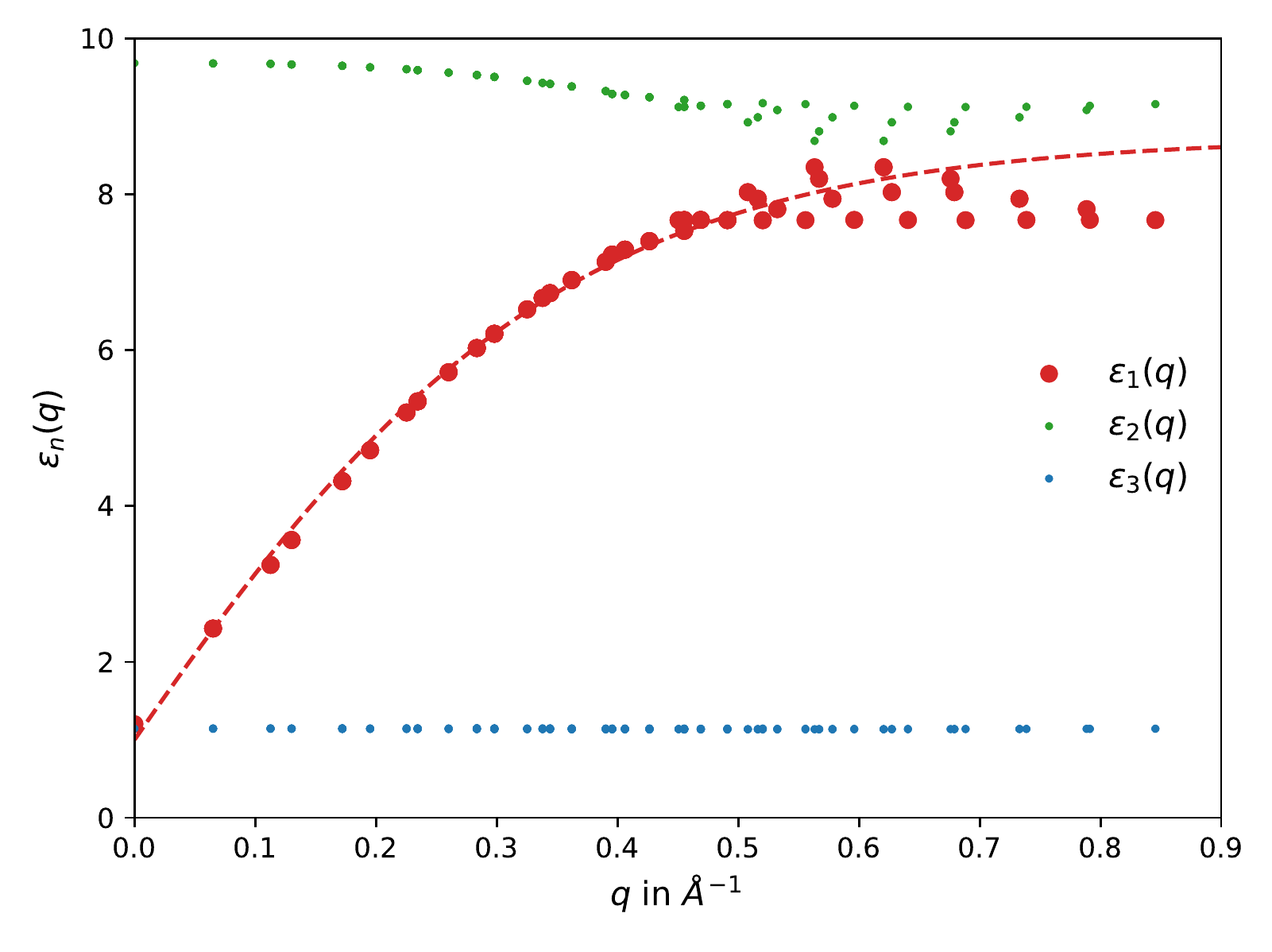}
         \caption{\textbf{Leading Screening Channels in the Wannier Function Continuum Electrostatics Approach.} Red, green, and blue dots correspond to $\varepsilon_1(q)$, $\varepsilon_2(q)$, and $\varepsilon_3(q)$, respectively. $\varepsilon_1(q)$ is fitted using the expression from Eq.~(\ref{eqn:WFCE}). The shown data is for the ($d$+$p$)-model. \label{fig:WFCE}}
        \end{figure}

        We start with the non-local bare Coulomb interaction of the CrI$_3$ monolayer as obtained from our cRPA calculations in momentum space. Within a matrix representation $v_{\alpha \beta}(q)$ using a product basis $\alpha, \beta = \{n,m\}$ we can easily diagonalize the Coulomb tensor
        \begin{align}
            v(q) = \sum_\nu v_\nu(q) \ket{v_\nu(q)} \bra{v_\nu(q)}
        \end{align}
        with $v_\nu(q)$ and $\ket{v_\nu(q)}$ being the corresponding eigenvalues and eigenvectors of the Coulomb tensors. Assuming that the eigenbasis does not drastically change upon the effects of the cRPA screening, we can thus represent the full cRPA Coulomb tensor as
        \begin{align}
            U(q) = \sum_\nu \frac{v_\nu(q)}{\varepsilon_\nu(q)} \ket{v_\nu(q)} \bra{v_\nu(q)},
        \end{align}
        where $\varepsilon_\nu(q)$ are the corresponding pseduo-eigenvalues of the dielectric tensor describing the different screening channels. In Fig.~\ref{fig:WFCE} we show the first three $\varepsilon_\nu(q)$ as a function of momentum $q$ and find that only one shows a significant dispersion, which we refer to as the ``leading'' eigenvalue in the following. This behaviour becomes clear upon investigating the corresponding eigenvectors in the basis of the two Cr atoms in the long-wavelength limit, i.e. for $q \rightarrow 0$:
        \begin{align}
            \ket{v_1(q)} = 
            \left| \begin{matrix} +1 \\ +1 \end{matrix} \right \rangle, \qquad
            \ket{v_2(q)} = 
            \left| \begin{matrix} +1 \\ -1 \end{matrix} \right \rangle.
        \end{align}
        The leading eigenvalue $v_1(q)$ thus renders Coulomb penalties for mono-pole-like perturbations (all orbitally resolved electronic densities on both Cr atoms are in phase), while the second eigenvalue $v_2(q)$ corresponds to Cr-Cr dipole-like Coulomb penalties. $\varepsilon_1(q)$ and $\varepsilon_2(q)$ thus correspondingly describe mono- and Cr-dipole-like screening. While the dipole-like screening from the environment is negligible, the mono-pole-like screening as rendered by $\varepsilon_1(q)$ is strongly affected. This classical electrostatic screening can be modeled by solving the Poisson equation for a dielectric slab of height $h$ embedded in some different dielectric environment \cite{RosnerWehling2015,jena_enhancement_2007,emelyanenko_effect_2008,keldysh_coulomb_1979} yielding
        \begin{align}
            \varepsilon_1(q) &= \frac{\varepsilon_1^{(0)} \left[ 1 - \tilde{\varepsilon}_0^{(1)} \tilde{\varepsilon}_0^{(2)} e^{-2q h} \right]}{1 + \left[ \tilde{\varepsilon}_0^{(1)} + \tilde{\varepsilon}_0^{(2)} \right] e^{-q h} + \tilde{\varepsilon}_0^{(1)} \tilde{\varepsilon}_0^{(2)} e^{-2q h}} \label{eqn:WFCE}
        \end{align}
        with
        \begin{align*}
    		\tilde{\varepsilon}_0^{(1)} = \frac{\varepsilon_1^{(0)} - \varepsilon_\text{sub}^\text{below}}{\varepsilon_1^{(0)} + \varepsilon_\text{sub}^\text{below}}, \qquad
    		\tilde{\varepsilon}_0^{(2)} = \frac{\varepsilon_1^{(0)} - \varepsilon_\text{sub}^\text{above}}{\varepsilon_1^{(0)} + \varepsilon_\text{sub}^\text{above}}.
        \end{align*}
        For $\varepsilon_\text{sub}^\text{above} = \varepsilon_\text{sub}^\text{below} = 1$ this describes the leading dielectric function of a free-standing monolayer, which we can fit perfectly to the cRPA data, as shown in Fig.~\ref{fig:WFCE} and yielding $h\approx5.2\,$\AA\ and $\varepsilon_1^{(0)} \approx 8.7$ (for the case of the ($d$+$p$)-model). With these fitting parameters we can now modify the full cRPA Coulomb tensor to describe environmental screening rendered by finite $\varepsilon_\text{sub}^\text{above}$ and $\varepsilon_\text{sub}^\text{below}$. Due to the monopole-like character of this environmental screening we only affect density-density Coulomb matrix elements in the very same way. Therefore Coulomb exchange elements, such as $J^H$ are not affected by the substrate screening. In the Supplemental Material we benchmark this approach to full cRPA calculations taking the screening from additional (strained) hBN layers into account.
        
\begin{acknowledgments}
    We thank M. Kaltak for sharing his cRPA implementation with us.
    MIK and ANR acknowledge support by European Research Council via Synergy Grant 854843 - FASTCORR. D.S. thanks financial support from EU through the MSCA project Nr. 796795 SOT-2DvdW. Part of this work was carried out on the Dutch national e-infrastructure with the support of SURF Cooperative.
\end{acknowledgments}

\appendix

\section{WFCE benchmark\label{sec:app:WFCE}}

\begin{figure}[htbp]
         \centering
         \includegraphics[width=0.49\textwidth]{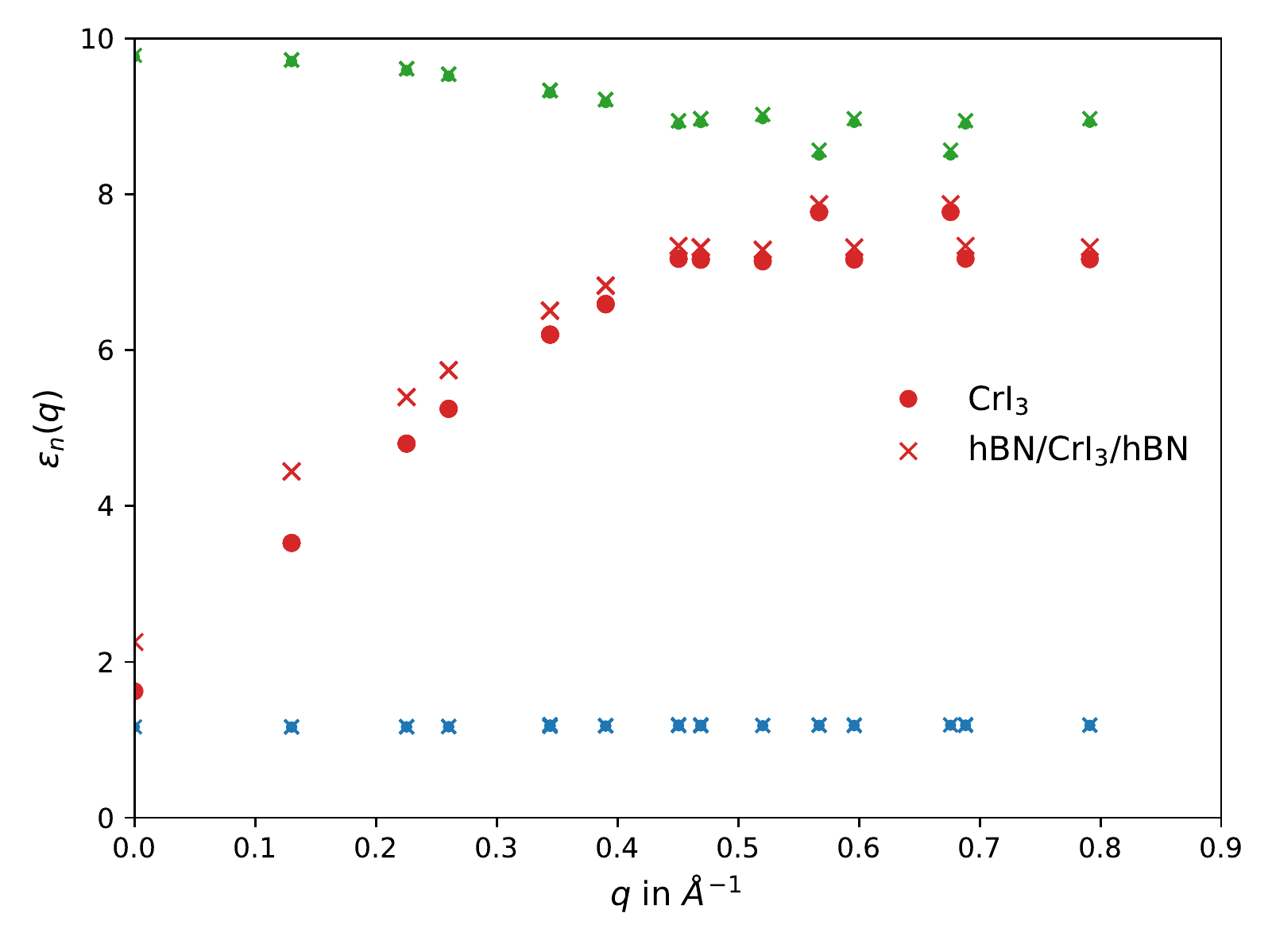}
         \caption{\textbf{Leading Screening Channels in the Wannier Function Continuum Electrostatics Approach for freestanding and hBN embedded CrI$_3$.} Dots represent the freestanding data and crosses the hBN embedded data. Red, green, and blue dots /crosses correspond to $\varepsilon_1(q)$, $\varepsilon_2(q)$, and $\varepsilon_3(q)$, respectively. The shown data is for the $d$-only model using an interlayer-distance of $h_{vac} = 20\,$\AA\ and $8 \times 8 \times 1$ $k$/$q$ meshes. \label{fig:app:WFCE}}
    \end{figure}

    To benchmark the validity of the WFCE approach for CrI$_3$ we calculated the local cRPA Coulomb tensors for freestanding CrI$_3$ and hBN-embedded CrI$_3$. For the latter we used a heterostructure consisting of a CrI$_3$ monolayer between two hBN monolayers. To keep the number of involved atoms minimal we strained the hBN lattice constant to $2.32\,$\AA\ (compared to $2.51\,$\AA) allowing us to use a primitive unitcell of CrI$_3$ embedded in $2 \times 2$ hBN monolayer supercells with $44$ atoms in total. The distance between the outer I atoms and the hBN layers was set to $4\,$\AA. For the cRPA calculations we used $8 \times 8 \times 1$ $k$ and $q$ meshes, $256$ ($384$) Kohn-Sham states for the freestanding (embedded) monolayer, and utilized the $d$-only model. The resulting leading screening channels are shown in Fig.~\ref{fig:app:WFCE} and the corresponding local Coulomb matrices are given in Tab.~\ref{tab:app:WFCE}. From Fig.~\ref{fig:app:WFCE} we see that indeed just the ``leading'' screening channel is affected by the dielectric environment formed by the hBN layers. From the Coulomb matrix elements in Tab.~\ref{tab:app:WFCE} we additionally see that indeed just the density-density elements are modified by the dielectric environment, while the Hund's exchange elements do not change. These are the two main properties rendering the WFCE method here applicable.
    
    \begin{table}
        \centering
        \caption{cRPA screened ($U$/$J$) local Coulomb interaction matrix elements from the $d$-only basis set for freestanding and hBN embedded CrI$_3$. Density-density matrix elements are listed on the upper triangle and Hund's exchange elements are underlined. \label{tab:app:WFCE}}
        
        \centering{CrI$_3$}
        \begin{ruledtabular}
            \begin{tabular}{l|lll|ll}
            $j / v$ & $t_{2g}$ & $t_{2g}$ & $t_{2g}$ & $e_{g}$ & $e_{g}$ \\ \hline 
            $t_{2g}$ & 3.40       &	2.40       &	2.40    &	2.52    &	2.40 \\
            $t_{2g}$ & \cg{0.48}  &	3.40       &	2.40    &	2.52    &	2.40 \\
            $t_{2g}$ & \cg{0.48}  &	\cg{0.48}  & 3.40       &	2.16    &	2.52\\ \hline
            $e_{g}$  & \cg{0.30}  &	\cg{0.30}  & \cg{0.46}  &  2.95     &	2.09\\
            $e_{g}$  & \cg{0.41}  &	\cg{0.41}  & \cg{0.25}  & \cg{0.43} &	2.95 
            \end{tabular}
        \end{ruledtabular}
        
        \centering{hBN/CrI$_3$/hBN}
        \begin{ruledtabular}
            \begin{tabular}{l|lll|ll}
            $J / U$ & $t_{2g}$ & $t_{2g}$ & $t_{2g}$ & $e_{g}$ & $e_{g}$ \\ \hline 
            $t_{2g}$ & 3.09       &	2.09       &	2.09    &	2.21    &	2.10 \\
            $t_{2g}$ & \cg{0.48}  &	3.09       &	2.09    &	2.21    &	2.10 \\
            $t_{2g}$ & \cg{0.48}  &	\cg{0.48}  & 3.09       &	1.85    &	2.33\\ \hline
            $e_{g}$  & \cg{0.30}  &	\cg{0.30}  & \cg{0.46}  &  2.64     &	1.78\\
            $e_{g}$  & \cg{0.41}  &	\cg{0.41}  & \cg{0.25}  & \cg{0.43} &	2.64 \\
            \end{tabular}
        \end{ruledtabular}
    \end{table}
\newpage
\section{Effective hoppings for the Kugel-Khomskii model}

    For the analysis of the Kugel-Khomskii model we need the effective hopping parameters $\tilde{t}^2$, which are listed in Tab.~\ref{tab:hoppings} and calculated via:
    \begin{align*}
        \tilde{t}^2_\text{t2g-t2g}(n.n.) &= \sum_{m,n \in t_{2g}} \left[ t^{Cr1-Cr2}_{mn}(R=0) \right]^2\\
        \tilde{t}^2_\text{t2g-eg}(n.n.) &= \sum_{m \in t_{2g}, n \in eg} \left[ t^{Cr1-Cr2}_{mn}(R=0) \right]^2\\ 
        \tilde{t}^2_\text{t2g-t2g}(n.n.n.) &= \sum_{m,n \in t_{2g}} \left[ t^{Cr1-Cr1}_{mn}(R=1) \right]^2\\
        \tilde{t}^2_\text{t2g-eg}(n.n.n.) &= \sum_{m \in t_{2g}, n \in eg} \left[ t^{Cr1-Cr1}_{mn}(R=1) \right]^2. \\
    \end{align*}
    For the ($d$+$p$)-model we use both, the direct Cr-Cr hoppings and those after integrating out the ligand $p$ contributions from the $R=0$ Wannier Hamiltonian. The latter enhances the effective hoppings, but the $\tilde{t}_\text{t2g-eg}$ channel is still smaller than the $\tilde{t}_\text{t2g-t2g}$ one.
  
    \begin{table}
            \centering
            \caption{Effective hoppings for the model exchange analysis with for the Kugel-Khomskii model in meV.}
            \label{tab:hoppings}
            \begin{ruledtabular}
                \begin{tabular}{lcccc}
                                 & $\tilde{t}_\text{t2g-t2g}$ & $\tilde{t}_\text{t2g-eg}$
                                 & $\tilde{t}_\text{t2g-t2g}'$ & $\tilde{t}_\text{t2g-eg}'$ \\
                    $d$-only     & $154$ & $145$        & $54$ & $64$ \\
                    $d$+$p$ (direct)      & $159$ & $29$         & $11$ & $10$ \\
                    $d$+$p$ ($p$ mediated) & $269$ & $119$ 
                \end{tabular}
            \end{ruledtabular}    
        \end{table}

\bibliography{biblio.bib}

\end{document}